\documentclass[twocolumn,secnumarabic,amssymb,nobibnotes,aps,prl,showpacs,showkeys]{revtex4-2}

\usepackage{graphicx}
\usepackage{amsmath}
\usepackage{hyperref}
\usepackage{footmisc}

\hypersetup{
	colorlinks=true,
	citecolor=blue,
	linkcolor=blue,
	filecolor=blue,      
	urlcolor=blue,}
\newcommand{\aof}{\mathring{a}_{\text{of}}^{(3)}}


\begin{document}	
	
\title{Investigation of the Lorentz invariance violation in two-neutrino double-beta decay} 
\author{S.A. Ghinescu$^{1,2,3}$, O. Ni\c{t}escu$^{1,2,3}$}
\author{S. Stoica$^{1}$}
\email[Corresponding author:]{sabin.stoica@cifra-c2unesco.ro}
\affiliation{
$^1$International Centre for Advanced Training and Research in Physics, PO Box MG12, 077125-M\u agurele, Romania \\
$^2$ "Horia Hulubei" National Institute of Physics and Nuclear Engineering, 30 Reactorului, POB MG-6, RO-077125, Bucharest-M\u agurele, Romania \\
$^3$ Faculty of Physics, University of Bucharest, 405 Atomi\c stilor, POB MG-11, RO-077125, Bucharest-M\u agurele, Romania}
\date{\today}

\begin{abstract}

We make a comprehensive investigation of the Lorentz invariance violation (LIV) effects that may occur in two-neutrino double-beta ($2\nu\beta\beta$) decay for all the experimentally interesting nuclei. We deduce the formulas for the LIV deviations and provide single and summed energy electron spectra and angular correlation between electrons with and without LIV contributions, to be used for constraining the LIV coefficient $\aof$. First, we confirm the shifting of the electron spectra to higher electron energies due to LIV for all nuclei. Next, we analyze other LIV signatures that can be used in LIV investigations. Thus, from the comparison of the electron and angular correlation spectra calculated with the inclusion of the LIV contributions, with their standard forms, information can be obtained about the strength versus observability of the LIV effects in the current experimental statistics. Then, we present the alternative method of constraining $\aof$  from the measurement of the angular correlation coefficient and estimate the statistics that different double-beta decay experiments should reach to constrain the LIV coefficient at the level of the current beta decay experiments. We hope that our work will improve the theoretical support and further stimulate the search for LIV in double-beta decay. 
\end{abstract}
%\pacs{ffff}
%\keywords{$2\nu\beta\beta$-decay, Lorentz invariance violation, angular corelation}
\maketitle
%\newpage
\setcounter{equation}{0}
\renewcommand{\theequation}{\arabic{equation}}
%\newpage 
%\section{Introduction} 
%\label{sec:intro} 
%\setcounter{equation}{0}
%\renewcommand{\theequation}{1.\arabic{equation}} 
\paragraph*{Introduction.} 
Investigation of the Lorentz invariance violation in $2\nu\beta\beta$ decay is an interesting research topic that is currently included in the study of this process. The theoretical framework underlying the estimation of the LIV effects in various physical processes is the Standard-Model extension (SME) theory, which incorporates Lorentz invariance violating operators of arbitrarily large dimension \cite{CK-PRD55,CK-PRD58,K-PRD69,KR-RMP2011}. Of particular interest is the minimal SME, where LIV effects can occur only through operators of mass dimension four or less \cite{KR-RMP2011}, which represents the theoretical background of many investigations, including those in the neutrino sector. The operators that couple to neutrinos can affect the neutrino oscillations, neutrinos velocity, and spectra of the electrons emitted in beta and double-beta decays \cite{KM-PRD69,Adam2012,Agnes-GALAXIES2021,Diaz-AHEP,Diaz-PRD89,DKL-PRD88}. 

Effects of LIV in the neutrino sector have been searched first in neutrino oscillation experiments such as Double-Chooz \cite{DC-PRD86}, MiniBoone \cite{MBoone-PLB718}, Ice Cube \cite{IC-PRD82}, MINOS \cite{Minos-PRD85}, SuperKamiokande \cite{SK-PRD91}, resulting in constraints of the LIV coefficients that control different couplings. However, according to the SME theory, the LIV effects in the neutrino sector can also be induced by the so-called oscillation-free operators of dimension three (countershaded effects), which do not affect the neutrino oscillations and hence can not be measured in these experiments. They are controlled by an oscillation-free (of) coefficient with four components, one time-like $\aof$ and three space-like $(a^{(3)}_{\rm of})_{1m}$, with $m=0, \pm 1$. Particularly, the LIV effects induced by the isotropic component of the countershaded operator can be searched in double-beta decay (DBD) experiments. This is because, in these experiments, the neutrinos are not measured, and only a global effect given by neutrinos of all orientations can be detectable \cite{Diaz-PRD88}. LIV signatures have recently been searched in DBD experiments such as  EXO \cite{EXO-200-PRD93}, GERDA \cite{GERDA-PhDThesis}, AURORA \cite{AURORA-2018}, NEMO-3 \cite{NEMO-3-2019,NEMO-3-PhdThesis}, CUORE \cite{CUORE-2019, CUORE-PhDThesis}, CUPID-0 \cite{CUPID-0-PRD100}, and the non-observation of the LIV effects resulted in constraints on the $\aof$ coefficient. These investigations were based until recently, on predictions of the electron spectra that were done with approximate (analytical) Fermi functions, built from electron wave functions (w.f.) obtained within a point-like nucleus model \cite{Primakoff-1959,Haxton-1984,Doi-1985,Suhonen-1998} and without screening effects. In two previous papers, we provided predictions of the single and summed energy electron spectra and angular correlation between electrons as well as their deviations due to LIV, calculated with improved electron w.f. \cite{NIT-2020, NIT-2021}. First, in Ref. \cite{NIT-2020} we compared the results of calculating $2\nu\beta\beta$ decay observables using Fermi functions obtained with different methods. We found that the differences in the values of the phase-space factors and decay rates calculated with different Fermi functions can be up to $30\%$. Thus, we concluded that the exact electronic w.f., obtained by numerically solving the Dirac equation in a realistic Coulomb-type potential, including the finite nuclear size correction and screening effects, are indicated for the accurate calculation of the phase space factors and further of the electron spectra and their LIV deviations. Next, using this method, we provided theoretical summed energy electron spectra for experimental LIV analyses for the $^{100}\text{Mo}$ nucleus. Then, in Ref. \cite{NIT-2021}, we extended the analysis of the LIV effects to single electron spectra and angular correlations between electrons. We discussed the LIV deviations that may occur in these spectra showing that they manifest differently for positive and negative values of the LIV coefficient $\aof$ and become more pronounced as the electron energy approaches the Q-value. We also proposed an alternative method to constrain $\aof$, namely through the measurement of the angular correlation coefficient. However, our analysis of the LIV effects in \cite{NIT-2021} was limited to $^{100}$Mo nucleus, for which the single state dominance (SSD) hypothesis (i.e., only the first $1^+$ state in the intermediate odd-odd nucleus contributes to the DBD rate \cite{Abad-1984,Simkovic_2001,Domin-2005}) can be used in calculations. 
  
In this paper, we extend the previous analyses to all nuclei that are currently being studied in DBD experiments, namely $^{48}$Ca, $^{76}$Ge, $^{82}$Se, $^{100}$Mo,  $^{110}$Pd, $^{116}$Cd, $^{130}$Te, $^{136}$Xe and $^{150}$Nd. We deduce the formulas for the LIV deviations and provide single electron spectra, summed energy electron spectra, and angular correlation between electrons calculated with and without LIV contributions, which are measured in $2\nu\beta\beta$ decay. Different from the $^{100}$Mo case, in most other studied nuclei, more $1^+$ states in the intermediate nucleus with higher energies can also contribute to the decay rate (HSD hypothesis). For these isotopes, the perturbation of the single electron spectra due to LIV may look different, as we will show. Next, we compare the electron and angular correlation spectra calculated with the inclusion of the LIV perturbations with their Standard Model (SM) forms and discuss the information that can be obtained about the strength versus observability of the LIV effects in the current experimental statistics. Finally, we present the alternative method of constraining $\aof$  from the measurement of the angular correlation coefficient and estimate the statistics that different double-beta decay experiments should reach to constrain this coefficient at the level of the current beta decay experiments.

\paragraph*{Theoretical formalism}
In this section we deduce the necessary formulas for the electron spectra, angular correlation, and angular correlation coefficient as well as for their perturbations due to Lorentz invariance violation. LIV effects in the neutrino sector can be estimated taking into account that the neutrino four-component momentum modifies from its standard expression $q^{\alpha} = (\omega, {\bf q})$  to  $q^{\alpha} = (\omega, {\bf q} + {\bf a}^{(3)}_{\rm of}-\mathring{a}^{(3)}_{\rm of} \bf \hat{q})$ \cite{KR-RMP2011,Diaz-PRD89,KT-PRL102}. In $2\nu\beta\beta$ decay this induces a change in the total decay rate that can be expressed as a sum of two terms \cite{Diaz-PRD89}: 
\begin{equation} 
	\Gamma_{\rm SME} = \Gamma_{\rm SM} + \delta \Gamma,
\end{equation}
where $\Gamma_{\rm SM}$ is the standard decay rate and $\delta \Gamma$ is the LIV contribution. 
The differential decay rate for the standard $2\nu\beta\beta$ decay process and for ground states (g.s) to g.s. transitions $0_{gs}^+\rightarrow0_{gs}^+$, can be expressed as  \cite{Haxton-1984,Doi-1985,Tomoda-1991,Kotila-2012}:
\begin{equation}
	d\Gamma_{\rm SM}=\left[\mathcal{A}+ \mathcal{B}\cos\theta_{12}\right]w_{\rm SM}d\omega_1d\varepsilon_1d\varepsilon_2d(\cos\theta_{12})
	\label{eq:DiferentialRate}
\end{equation}
where $\varepsilon_{1,2}$ are the electron energies, $\omega_{1,2}$ are the antineutrino energies, and $\theta_{12}$ is the angle between the directions of the two emitted electrons. In what follows, we adopt the natural units ($\hbar=c=1$). Within the SM framework, the term $w_{\rm SM}$ is given by:
\begin{equation}
	w_{\text{SM}}=\frac{g_A^4G_F^4\left|V_{ud}\right|^4}{64\pi^7}\omega_1^2\omega_2^2p_1p_2\varepsilon_1\varepsilon_2
\end{equation} 
where $g_A$ is the axial vector constant, $G_F$ is the Fermi coupling constant, $V_{ud}$ is the first element of the Cabibbo-Kobayashi-Maskawa matrix and $p_{1,2}$ are the momenta of the electrons.

The $\mathcal{A}$ and $\mathcal{B}$ quantities are products of nuclear matrix elements (NMEs)  and phase-space factors (PSFs) for the $2\nu\beta\beta$ decay mode. Their explicit expressions can be found in many papers on DBD (see  for example  \cite{Doi-1985, Tomoda-1991, NIT-2021}). 

After the integration over the lepton energies, the derivative of the decay rate with respect to the cosine of the angle $\theta_{12}$ can be written as a sum between the spectrum part and  angular correlation part:

\begin{equation}
\label{SMDecayRate}
	\frac{d\Gamma_{\rm SM}}{d(\cos\theta_{12})}=\frac{1}{2}\left(\Gamma_{\rm SM} + \Lambda_{\rm SM} \cos\theta_{12} \right) = \frac{1}{2}\Gamma_{\rm SM}\left(1+ \kappa_{\rm SM} \cos\theta_{12}\right).
\end{equation}
Here, $\kappa_{\rm SM} = \Lambda_{\mathrm{SM}}/\Gamma_{\mathrm{SM}}$ is the angular correlation coefficient.  $\Lambda_{\mathrm{SM}}$, the angular part of the decay rate, is also affected by LIV and, like the spectrum part, can be  written as a sum between its SM form and the LIV deviation:
\begin{equation}
 \Lambda_{\rm SME}=\Lambda_{\rm SM}+\delta\Lambda.
\end{equation} 
We note that after integration over $\cos\theta_{12}$ only the spectrum part gives contribution to the total DBD decay rate. Using the closure approximation, the $2\nu\beta\beta$ decay rate can be expressed in a factorized form \cite{Haxton-1984,Doi-1985,Tomoda-1991}:  

\begin{align}
	\begin{aligned}
		\frac{\Gamma}{\ln 2}&=g_A^4\left|M\right|^2G, \\
		\frac{\Lambda}{\ln 2}&=g_A^4\left|M\right|^2H,
		\label{eq:decayrate_factorization}
	\end{aligned}
\end{align}
\noindent
where $M$ are NMEs which depend on the nuclear structure of the nuclei involved in the decay, and $G$ and $H$ are PSFs which include the distortion of the electrons w.f. by the Coulomb field of the daughter nucleus. Since we refer to the LIV effects induced by the neutrino behavior, only PSFs are subject to the LIV modifications, namely:
\begin{eqnarray}
G_{\rm SME}=G_{\rm SM}+\delta G, \\
H_{\rm SME}=H_{\rm SM}+\delta H
\end{eqnarray}

The phase-space factors for the $2\nu\beta\beta$ transitions to final ground states can be written in a compact form as follows \cite{NIT-2021}:
\begin{widetext}
	\begin{eqnarray}
	\begin{Bmatrix}
	\label{PSF1}
	G_{\rm SM}\\
	\delta G
	\end{Bmatrix}=&&
	\frac{\tilde{A}^2G_F^2\left|V_{\text{ud}}\right|^2m_e^9}{96\pi^7\ln2}\frac{1}{m_e^{11}}\int_{m_e}^{E_I-E_F-m_e}d\varepsilon_{1}\varepsilon_{1}p_1\int_{m_e}^{E_I-E_F-\varepsilon_{1}}d\varepsilon_{2}\varepsilon_{2}p_2\nonumber\\
	&&\times\int_{0}^{E_I-E_F-\varepsilon_{1}-\varepsilon_{2}}d\omega_1\omega_2^2a(\varepsilon_{1},\varepsilon_{2})\left[\langle K_N\rangle^2+\langle L_N\rangle^2+\langle K_N\rangle\langle L_N\rangle\right]\begin{Bmatrix}
	\omega_1^2\\
	4\mathring{a}^{(3)}_{\rm of}\omega_1
	\end{Bmatrix} \\
	\begin{Bmatrix}
	\label{PSF2}
	H_{\rm SM}\\
	\delta H
	\end{Bmatrix}=&&
	\frac{\tilde{A}^2G_F^2\left|V_{\text{ud}}\right|^2m_e^9}{96\pi^7\ln2}\frac{1}{m_e^{11}}\int_{m_e}^{E_I-E_F-m_e}d\varepsilon_{1}\varepsilon_{1}p_1\int_{m_e}^{E_I-E_F-\varepsilon_{1}}d\varepsilon_{2}\varepsilon_{2}p_2\nonumber\\
	&&\times\int_{0}^{E_I-E_F-\varepsilon_{1}-\varepsilon_{2}}d\omega_1\omega_2^2b(\varepsilon_{1},\varepsilon_{2})\left[\frac{2}{3}\langle K_N\rangle^2+\frac{2}{3}\langle L_N\rangle^2+\frac{5}{3}\langle K_N\rangle\langle L_N\rangle\right]\begin{Bmatrix}
	\omega_1^2\\
	4\mathring{a}^{(3)}_{\rm of}\omega_1
	\end{Bmatrix} ,
	\end{eqnarray}
\end{widetext}
where $m_e$ is the electron mass.

The quantities $\langle K_N\rangle$ and  $\langle L_N\rangle$ are kinematic factors that depend on the lepton energies ($\epsilon$, $\omega$), the g.s. energy of the parent nucleus ($E_I$), and an averaged energy of the excited $1^+$ states in the intermediate nucleus ($\langle E_N \rangle$). Replacing the energies of the $1^+$ states with an average energy is called the closure approximation and allows to express the $2\nu\beta\beta$ decay rate as a product of the PSF and NME parts (see Eq. ~\ref{eq:decayrate_factorization}). The expressions of the kinematic factors $\langle K_N \rangle$ and $\langle L_N \rangle$ are given in many papers about the double-beta decay topic (see for example \cite{Haxton-1984}):
	\begin{align}
	\begin{aligned}
	\label{eq:KnDef}
	\langle K_N\rangle=
	{1\over \varepsilon_1+\omega_1+\langle E_N\rangle-E_I}+
	{1\over \varepsilon_2+\omega_2+\langle E_N\rangle-E_I}\\
	\langle L_N\rangle=
	{1\over \varepsilon_1+\omega_2+\langle E_N\rangle-E_I}+
	{1\over \varepsilon_2+\omega_1+\langle E_N\rangle-E_I}.
	\end{aligned}
	\end{align}
The energy $\langle E_N\rangle-E_I$ is determined from the approximation
$\tilde{A}=[W_0/2+\langle E_N\rangle -E_I]$, where
$\tilde{A}=1.12A^{1/2}$ (in MeV) gives the energy of the giant Gamow-Teller resonance in the intermediate nucleus and $W_0=E_I-E_F$, where $E_F$ is the g.s. energy of the daughter nucleus. We note that in many calculations, simplified expressions of these factors are used, namely: $\langle K_{N}\rangle \simeq \langle L_N \rangle \simeq 2/\tilde{A}$. With this approximation, the PSF formulas and their LIV deviations simplify much, but some accuracy is lost as well \cite{NIT-2020}.

To provide good predictions for the single and summed energy electron spectra, angular correlation between electrons, as well as for their deviations due to LIV, accurate calculations of the $G_{\text{SM}}$ and $H_{\text{SM}}$ phase space factors and their deviations are required. This implies accurate calculations of the integrals in Eqs.~\ref{PSF1} and \ref{PSF2} which contain the factors $a(\epsilon_1,\epsilon_2)$ and $b(\epsilon_1,\epsilon_2)$. These quantities are built with electron w. f. obtained by solving the Dirac equation in a realistic Coulomb-type potential, including the finite nuclear size (FNS) and screening effects. 

The functions $a(\varepsilon_1,\varepsilon_2)$ and $b(\varepsilon_1,\varepsilon_2)$ are defined as \cite{Kotila-2012,NIT-2020}
\begin{align}
\label{KNLN}
\begin{aligned}
&a(\varepsilon_1,\varepsilon_2)=\left|\alpha^{-1-1}\right|^2+\left|\alpha_{11}\right|^2+\left|\alpha_{1}^{\hspace{0.16cm}-1}\right|^2+\left|\alpha^{-1}_{\hspace{0.35cm}1}\right|^2\\
&b(\varepsilon_1,\varepsilon_2)=-2\Re\{\alpha^{-1-1}\alpha_{11}^*+\alpha^{-1}_{\hspace{0.35cm}1}\alpha_{1}^{\hspace{0.16cm}-1*}\}
\end{aligned}
\end{align}
with
\begin{align}
\label{eq:WavefunctionsProducts}
\begin{aligned}
\alpha^{-1-1}&=g_{-1}(\varepsilon_1)g_{-1}(\varepsilon_2),
\alpha_{11} = f_{1}(\varepsilon_1)f_{1}(\varepsilon_2),\\
\alpha_{1}^{\hspace{0.16cm}-1}&=f_{1}(\varepsilon_1)g_{-1}(\varepsilon_2),
\alpha^{-1}_{\hspace{0.35cm}1}= g_{-1}(\varepsilon_1)f_{1}(\varepsilon_2).
\end{aligned}
\end{align}
\noindent
where $f_{1}(\varepsilon_1)$ and $g_{-1}(\varepsilon_2)$ are the electron radial wave functions evaluated on the surface of the daughter nucleus:
\begin{align}
\begin{aligned}
g_{-1}(\varepsilon)&=\int_{0}^{\infty}g_{-1}(\varepsilon,r)\delta(r-R)dr\\
f_{1}(\varepsilon)&=\int_{0}^{\infty}f_{1}(\varepsilon,r)\delta(r-R)dr,
\end{aligned}
\end{align}
where $R=r_0A^{1/3}$, $r_0=1.2$ fm.

In the PSF evaluation for LIV analyses, we included the full expressions of $\langle K_N \rangle$ and $\langle L_N \rangle$ from Eq.~\ref{KNLN}, while in previous calculations, their simplified expressions mentioned above are used. Our method of calculation and the comparison of the results with other methods are described in detail in \cite{NIT-2020}, where we showed that using exact electron w.f. instead of approximative ones is more reliable in calculating the PSF values.

By differentiating the $2\nu\beta\beta$ decay rate expression versus the total energy of one electron ($\varepsilon_1$), one gets the single electron spectrum \cite{Doi-1985,Tomoda-1991,Kotila-2012}:
\begin{equation}\label{eq:SingleElectronSpectra}
	\frac{d\Gamma_{\rm SME}}{d\varepsilon_1} = C\frac{dG_{\rm SM}}{d \varepsilon_1}.
\end{equation}
Similarly, one gets the  summed energy spectrum of the two electrons:
\begin{equation}\label{eq:SumElectronSpectra}
	\frac{d\Gamma_{\rm SME}}{dK} = C\frac{dG_{\rm SM}}{d K}
\end{equation}
where $ K\equiv \varepsilon_1 + \varepsilon_2 -2m_e $ is the total kinetic energy of the two electrons. $C$ is a constant including the nuclear matrix elements. 

Also, by differentiating the decay rate versus $\varepsilon_1$ and $\cos\theta_{12}$, one gets the angular correlation, $\alpha_{\text{SM}}$, between the two emitted electrons:
\begin{align}
	\label{eq:DiffDecayRate_SME}
	\begin{aligned}
		&\frac{d\Gamma_\text{SM}}{d \varepsilon_1 d(\cos\theta_{12})}=C\frac{d G_{\rm SM}}{d\varepsilon_1} \left[1+\alpha_{\text{SM}}\cos\theta_{12}\right].
	\end{aligned}
\end{align}
where $ \alpha_{\text{SM}} \equiv (dH_{\rm SM}/d\varepsilon_1)/(dG_{\rm SM}/d\varepsilon_1)$ is the SM angular correlation.

In \cite{NIT-2021}, we calculated the expressions of these quantities and their LIV deviations for the single electron spectrum:

\begin{equation}\label{eq:SingleElectronSpectra}
\frac{d\Gamma_{\rm SME}}{d\varepsilon_1} = C\frac{dG_{\rm SM}}{d \varepsilon_1}\left(1+\aof \chi^{(1)}(\varepsilon_1)\right),
\end{equation}
and summed energy electron spectrum:
\begin{equation}\label{eq:SumElectronSpectra}
\frac{d\Gamma_{\rm SME}}{dK} = C\frac{dG_{\rm SM}}{d K}\left(1+\aof \chi^{(+)}(K)\right).
\end{equation}
Here,
\begin{equation}\label{eq:LIVdeviations}
\chi^{(1)}(\varepsilon_1) = \frac{d(\delta G)}{d \varepsilon_1}/\frac{dG_{\rm SM}}{d \varepsilon_1}
\end{equation}
and
\begin{equation}
\chi^{(+)}(K) = \frac{d(\delta G)}{d K}/\frac{dG_{\rm SM}}{d K}
\end{equation}
are quantities that incorporate the deviations of the electron spectra from their standard (SM) forms. 

The relation between the LIV-perturbed angular correlation and its standard form can be extracted from the expression of the derivative of the decay rate versus the total energy of an electron and the $\cos\theta_{12}$:
\begin{align}
\label{eq:DiffDecayRate_SME}
\begin{aligned}
&\frac{d\Gamma_\text{SME}}{d \varepsilon_1 d(\cos\theta_{12})}=C\frac{d G_{\mathrm{SM}}}{d\varepsilon_1}\times\\
&\left[1+\aof \chi^{(1)}(\epsilon_1)+\left(\alpha_{\text{SM}}+\aof\frac{d(\delta H)/d\varepsilon_1}{dG_{\rm SM}/d\varepsilon_1}\right)\cos\theta_{12}\right].
\end{aligned}
\end{align}
with
\begin{equation}
\label{eq:alpha_sme}
\alpha_{\text{SME}} = \alpha_{\text{SM}} + \aof \frac{d(\delta H)/d\varepsilon_1}{dG_{\rm SM}/d\varepsilon_1}
\end{equation}

Differentiating the decay rate expression versus $\cos\theta_{12}$
\begin{align}
\label{eq:k_sme}
\begin{aligned}
&\frac{d\Gamma_\text{SME}}{d(\cos\theta_{12})}=CG_{\rm SM}\times\\
&\left[1+\aof\frac{\delta G}{G_{\rm SM}}+\left(\kappa_{\text{SM}}+\aof\frac{\delta H}{G_{\rm SM}}\right)\cos\theta_{12}\right],
\end{aligned}
\end{align}
we can identify (in round brackets) the SME expression of the angular correlation coefficient $\kappa_{\text{SME}}$ and the relation with its standard form. For an independent treatment with respect to $\aof$, we define $\xi_{\text{LIV}} \equiv \delta H/G_{\mathrm{SM}}$ in units of $\mathrm{MeV}^{-1}$. 
Finally, the LIV-perturbed angular correlation coefficient can also be written as,
\begin{equation}
	\label{LIVAngularCorrelationFactor}
	\kappa_{\rm SME}=\frac{\Lambda_{\rm SM}}{\Gamma_{\rm SM}}+\frac{\delta \Lambda}{\Gamma_{\rm SM}}.
\end{equation}
The first term in the r.h.s of Eq.~\ref{LIVAngularCorrelationFactor} is the standard angular correlation coefficient, $\kappa_{\rm SM}$, and the second one is its LIV deviation. 

\paragraph*{Results and discussions.}

\begin{figure*}
	\centering{
		\includegraphics[width=0.8\textwidth]{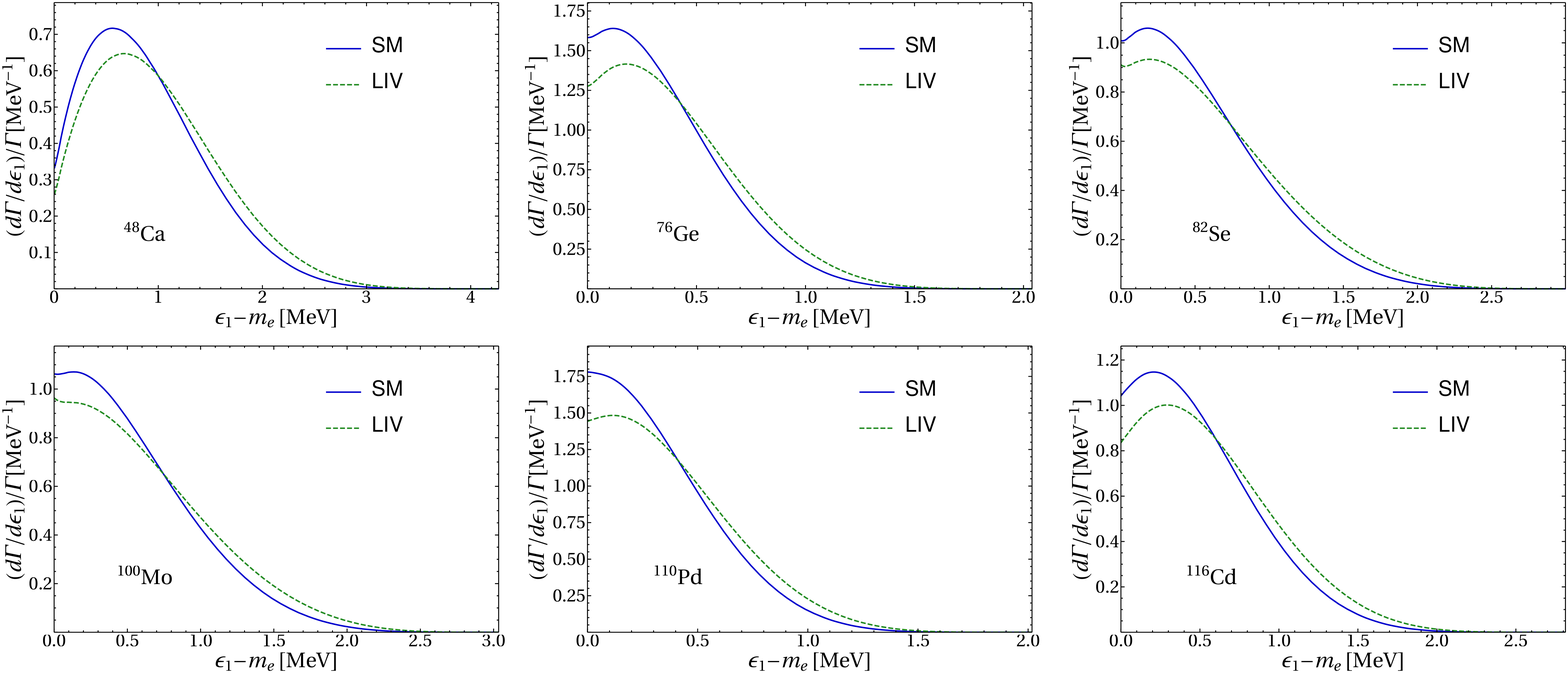}
		\includegraphics[width=0.52\textwidth]{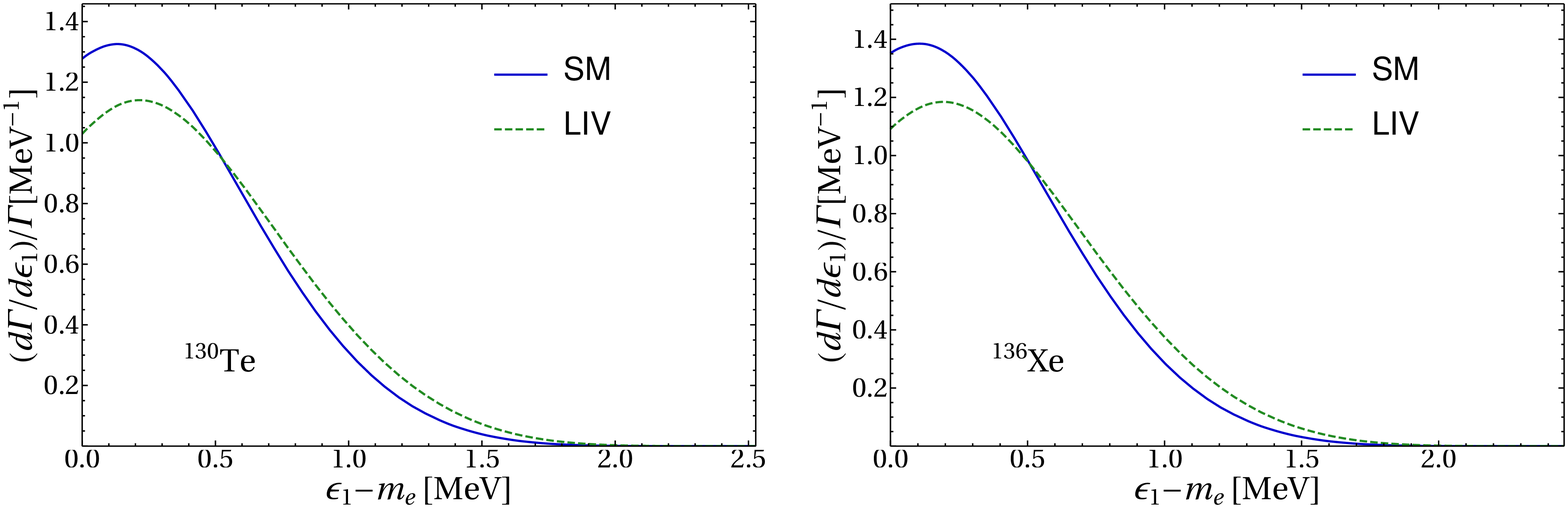}
	}
	\caption{(Color online) Normalized $2\nu\beta\beta$ single electron spectra within SM with solid line and the first order contribution in $\aof$ due to LIV with dashed line. See text for the assumption on the hypothesis used.  }
	\label{fig:singlespectra}
\end{figure*}

We calculate the single and summed energy electron spectra, angular correlation spectra and angular correlation coefficient, along with their LIV deviations from the standard forms for all nuclei that are investigated in DBD experiments, i.e. $^{48}$Ca, $^{76}$Ge, $^{82}$Se, $^{100}$Mo, $^{116}$Cd, $^{130}$Te, $^{136}$Xe and $^{150}$Nd. As already mentioned, we use electron radial wave functions obtained as solutions of the Dirac equation in a Coulomb potential that encodes the finite-size and the atomic screening of the final nucleus. We numerically solve the radial Dirac equation with the subroutine package RADIAL \cite{Salvat-1991,Salvat-CPC2019}. Following this procedure, the truncation errors are completely avoided, and the radial wave functions are obtained with the desired accuracy. Thus, the numerical solutions can be considered as exact for the given input potential. More details about the electrostatic potential that we use can be found in Refs. \cite{SM-2013,MPS-2015,NIT-2020}. In calculations, we use either the SSD or HSD hypothesis as follows. The SSD hypothesis has been experimentally validated for the $^{82}\mathrm{Se}$ \cite{CUPID-0-PRL2019} and $^{100}\mathrm{Mo}$ \cite{NEMO-3-2019} nuclei and we used it for these isotopes. This means we replaced $\langle E_{N} \rangle$ in the formulas from the previous section, with the energy of the first $1^+$ intermediate state ($E_{1_1^+}$). For $^{150}\mathrm{Nd}$ nucleus, the dominant DBD transition also occurs through the first $1^+$ state in the intermediate nucleus, $^{150}\mathrm{Pm}$, but transitions through other $1^+$  states of higher energies, also contribute and must be included in the calculation so that the DBD rate value is reproduced \cite{150Nd-PRC2011}. Thus, we calculated the single electron spectra using both (SSD and HSD) hypotheses for this nucleus. In Fig. \ref{fig:singlespectra} we present the normalized standard and LIV-perturbed spectra for all nuclei except $^{150}\mathrm{Nd}$. For the nuclei where the SSD hypothesis applies, we used the following values for the $1+$ state energies ($E_{1_1^+}-E_I$): $-0.338$ MeV for $^{82}\mathrm{Se}$, $-0.343$ MeV for $^{100}\mathrm{Mo}$ and $-0.315$ MeV for $^{150}\mathrm{Nd}$.  As can be seen, the main difference between the calculations for different nuclei can occur at low electron energies. For the nuclei where the HSD hypothesis applies, the LIV spectra increase first monotonously with increasing energy and reach their maxima at energies not close to 0. On the other hand, for the isotopes where the SSD hypothesis applies, the LIV spectra (except $^{82}Se$) show a local maximum at $\varepsilon_{1}\to0$. For more precise information, in Table I, we give the position of global maxima of the LIV spectra for all nuclei. Concluding, regardless of the hypothesis assumed, the overall effect of LIV  on the single electron spectra in all nuclei is a shift of the spectra towards higher electron energies, as shown in Ref. \cite{NIT-2021} in the case of $^{100}\text{Mo}$. This is an effect similar to that found in the summed energy electron spectra \cite{NIT-2020}.

\begin{figure*}
	\includegraphics[width=0.8\textwidth]{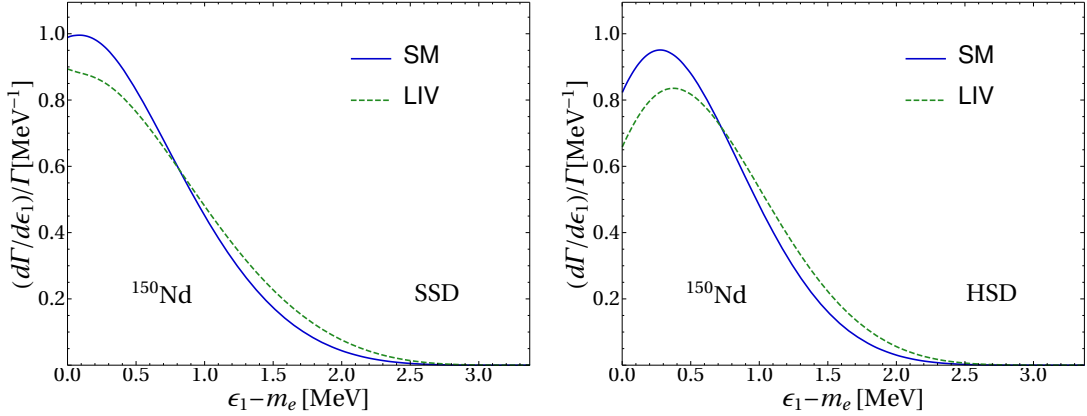}
	\caption{(Color online) Normalized single electron spectra within SM with solid line and the first order contribution in $\aof$ due to LIV with dashed line for $2\nu\beta\beta$ decay of $^{150}\mathrm{Nd}$. We assumed the SSD hypothesis in the left panel and HSD in the right panel.  }
	\label{fig:150NdSSDvsHSD}	
\end{figure*}

\begin{figure*}
	\includegraphics[width=0.8\textwidth]{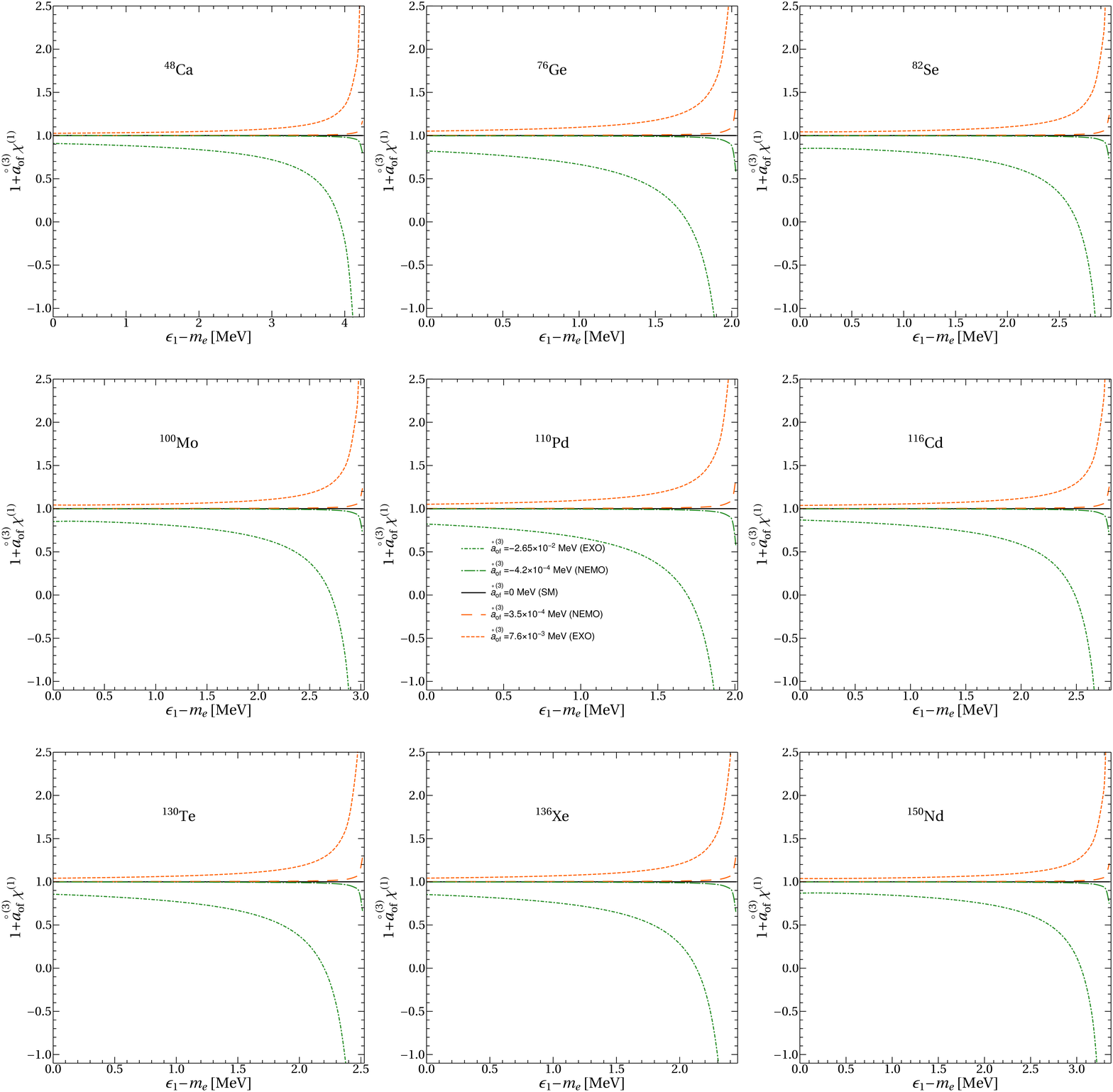}
	\caption{(Color online) The quantity $\chi^{(1)}(\varepsilon_1)$ depicted for current limits of $\aof$ (dashed for upper limit and dot-dashed for lower limit). The solid line at $\chi^{(1)}(\epsilon_1)= 0$ represents the SM prediction.  }
	\label{fig:Chi1allnuclei}
\end{figure*} 

\begin{figure*}
	\includegraphics[width=0.8\textwidth]{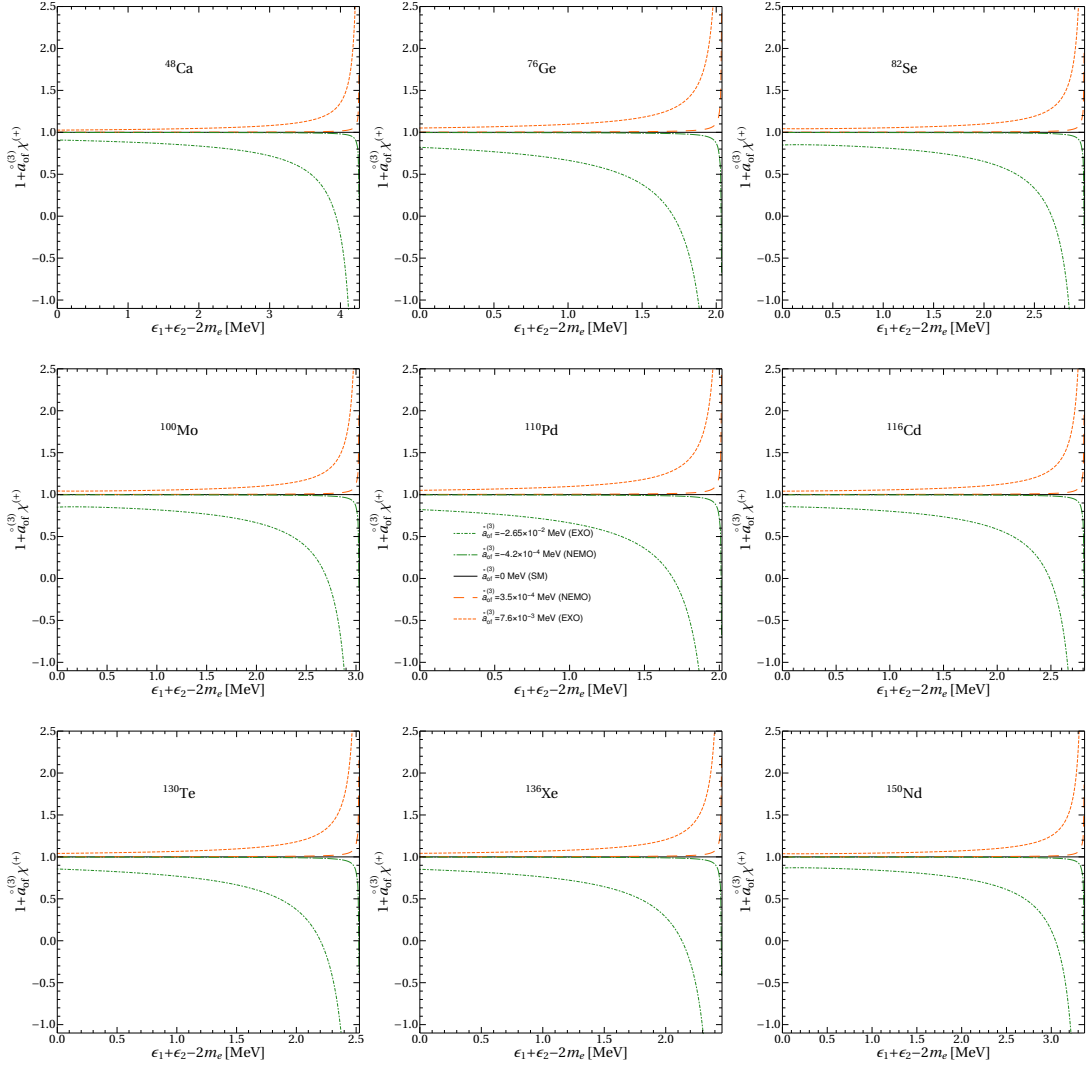}
	\caption{(Color online)The quantity $\chi^{(+)}(K)$ depicted for current limits of $\aof$.The same conventions as in Fig.~\ref{fig:Chi1allnuclei} are used.}
	\label{fig:ChiSumAllnuclei}
\end{figure*}

Next, we discuss other LIV signatures resulting from the comparison of the single and summed energy electron spectra and angular correlation perturbed by LIV with their standard forms. We note first that in the previous works \cite{Diaz-PRD89, EXO-200-PRD93, NEMO-3-2019, NIT-2020} the LIV effects were presented by plotting separately, on the same figure, the normalized summed energy electron spectra calculated within SM, and their LIV deviations. Thus, as we already mentioned, it was concluded that the LIV effects (if they exist) manifest as a global shift of the electron spectra to higher electron energies. Further, using the theoretical predictions for the summed energy electron spectra and from the non-observation of such deviations, constraints on the LIV $\aof$ coefficient are deduced. Several DBD experiments reported such limits \cite{EXO-200-PRD93, GERDA-PhDThesis, AURORA-2018, CUPID-0-PRD100,NEMO-3-2019, NEMO-3-PhdThesis}. Besides the analyzes on the summed energy electron spectra reported in these references, we presented in \cite{NIT-2021} another analysis of the LIV signatures by comparing the electron spectra (single and summed energy) and angular correlation calculated with and without LIV contributions. This was done for the $^{100}\mathrm{Mo}$ nucleus for which the SSD hypothesis holds. Here, we extend this analysis to all nuclei. 
Thus, in Fig.~\ref{fig:Chi1allnuclei} we plot the quantity $1 + \aof \chi^{(1)}(\varepsilon_1)$, which represents the ratio between the single electron spectrum calculated with the LIV contributions and its standard forms for all nuclei. The calculations are performed with  two (extreme) sets of $\aof$ limits, namely those reported by the EXO collaboration $-2.65\times 10^{-2} \mathrm{MeV} \le \aof \le 7.6\times 10^{-3} \mathrm{MeV}$\cite{EXO-200-PRD93} and those reported by the NEMO-3 collaboration $-4.2\times 10^{-4} \mathrm{MeV} \le \aof \le 3.5\times 10^{-4}\mathrm{MeV}$\cite{NEMO-3-2019}. Other limits reported until now can be found in \cite{KR-ARXIV, KR-RMP2011}. The horizontal line equal to $1$ represents the ratios in the absence of LIV effects, while the curves situated over or under this line represent the deviations when the LIV corrections are included. The position of the curves is dictated by the sign of the $\aof$ coefficient, over the horizontal unity line for positive values of $\aof$ and under this line for negatives values of this coefficient. As we mentioned in \cite{NIT-2021}, the increased divergences between the standard and the LIV perturbed spectra are due to a slower descent (in absolute value) of the LIV spectrum with respect to the standard one at the end of the energy interval (near $Q$-value). As seen, for $\aof$ limits reported by \cite{EXO-200-PRD93} the deviations of the single electron spectra due to LIV are quite pronounced (even for electron energies much lower than the  $Q$-value), and they should have been seen already, which did not happen. For more stringent limits of $\aof$, as those reported by NEMO-3 \cite{NEMO-3-2019}, the deviations are very small and cannot be seen in the current experimental statistics. However, in future DBD experiments, such as the SuperNEMO experiment, which targets $10^3$ times the statistics from NEMO-3 for $^{100}$Mo, these LIV deviations might be observed. These observations are valid for all the studied nuclei.
However, a drawback of the single electron spectra is that they can only be measured in DBD experiments with electron tracking systems. That is why we present a similar analysis for the summed energy electron spectra that are measured in all the DBD experiments and with higher statistics than the single electron spectra. In Fig.~\ref{fig:ChiSumAllnuclei}, we plot the ratio between the summed energy spectra of electrons calculated with the LIV contributions and their standard forms. One can see LIV effects with similar shapes, as in the case of the single electron spectra, and the same arguments are valid to explain them. From the analysis of the deviations of these predicted electron spectra, estimations on the magnitude and observability of the LIV effects in the different statistics can be made. 

\begin{figure*}
	\includegraphics[width=0.8\textwidth]{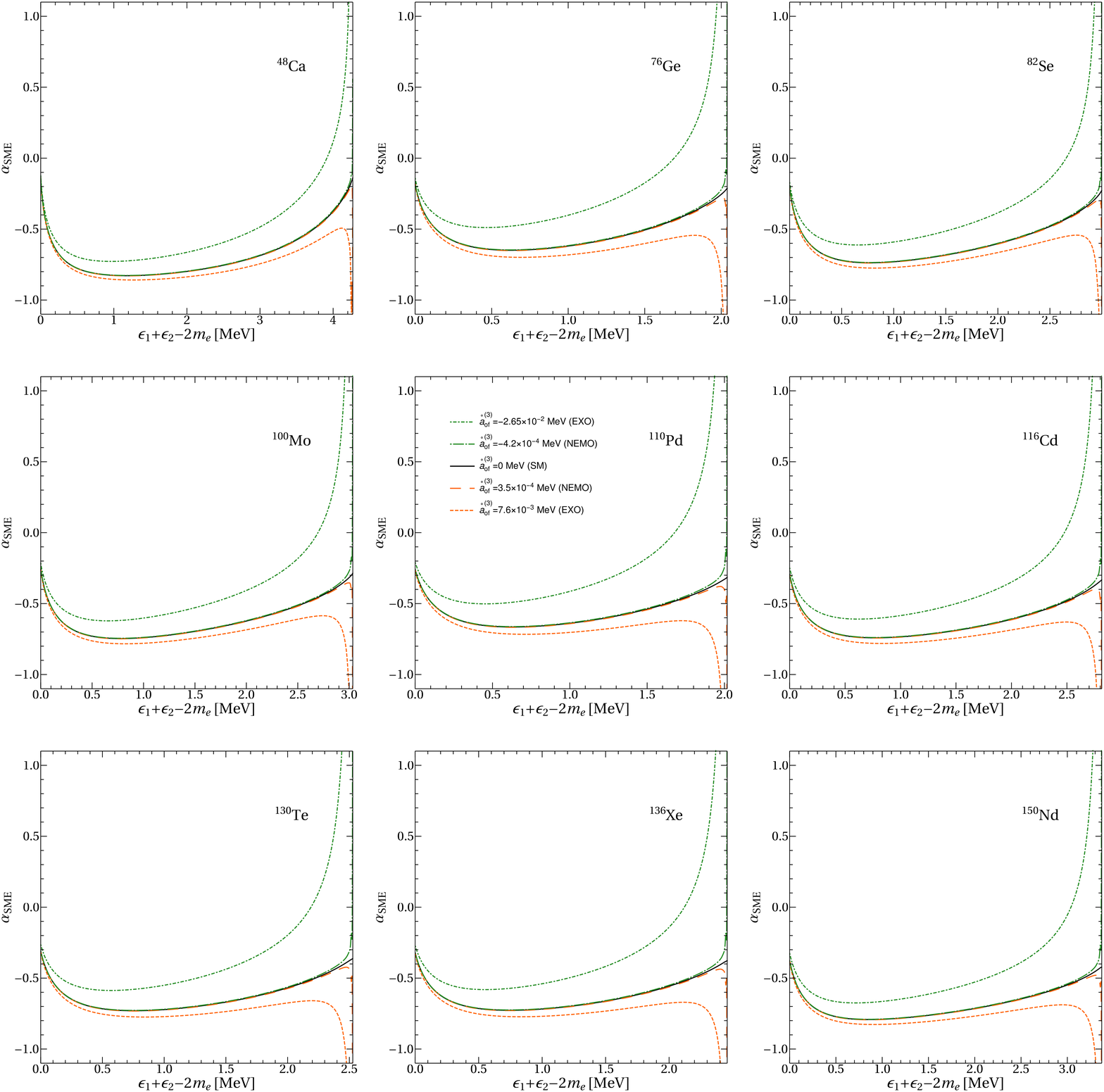}
	\caption{(Color online) The angular correlation spectrum plotted for the current limits of $\aof$. The same conventions as in Fig.~\ref{fig:Chi1allnuclei} are used.}
	\label{fig:angcorr_allnuclei}
\end{figure*}

Further, we discuss the LIV effects on the angular correlation $\alpha$ and the value of the angular correlation coefficient $k$. In Fig.~\ref{fig:angcorr_allnuclei} the angular correlation spectra for all the nuclei are plotted with the same conventions as in Fig.~\ref{fig:ChiSumAllnuclei}. As seen, deviations of the angular correlation curves from their standard forms may manifest even at low electron energies, and they increase much in the vicinity of the $Q$-value for the $\aof$ values reported by EXO. Again,  for the $\aof$ values reported by NEMO3, these deviations cannot be seen in the current experimental statistics. We also note that distinctively from the electron spectra, the total angular correlation spectrum exceeds the standard spectrum for negative values of $\aof$ because $\delta H$ is also negative, making the LIV contribution positive (see Eq. 24). Regarding the theoretical electron and angular correlation spectra discussed above, we mention that we can provide upon request detailed numerical predictions of these spectra to be used in DBD experiments for the LIV investigation.

Finally, we refer to the angular correlation coefficient, $k$, defined in Eqs.~\ref{SMDecayRate} and \ref{LIVAngularCorrelationFactor} of the previous section. As shown in Ref. \cite{NIT-2021} it can also be used to constrain the $\aof$ coefficient and estimate quickly (albeit grossly) the number of the $2\nu\beta\beta$ events needed to put a certain limit on $\aof$. $k_{\text{SME}}$ can be determined in the DBD experiments with electron tracking systems by using the forward-backward asymmetry \cite{Arnold-2010},
		
\begin{align}
A\equiv\frac{\int^0_{-1}\frac{d\Gamma}{dx}dx-\int^1_{0}\frac{d\Gamma}{dx}dx}{\Gamma}
=\frac{N_{+}-N_{-}}{N_{+}+N_{-}}=\frac{1}{2}k_{\mathrm{SME}},
\end{align}
where $x=\cos\theta_{12}$ and $N_{-}(N_{+})$ are the $2\nu\beta\beta$ events with the angle $\theta_{12}$ smaller (larger) than $\pi/2$. Assuming that the experimental value of this coefficient is compatible at 90\% CL with the SM value and considering only statistical uncertainties in the number of events recorded, one can compute the number of events needed to constrain the $\aof$ coefficient at a specific value. In Table~\ref{tab:KXiNevts}, we give the values of $k_{\mathrm{SM}}$ and $\xi_{\mathrm{LIV}}$ computed as described in the previous section. In the last column, we also give the number of events needed to constrain the upper limit of $\aof$ at the same value obtained from the tritium decay (i.e., $|\aof| < 3\times 10^{-5}\mathrm{MeV}$ \cite{KR-ARXIV}). We also indicate the nuclei for which we have employed the SSD hypothesis by subscript. In these cases, the $\tilde{A}$ value has been taken from \cite{Kotila-2012}. The rest of the nuclei have been treated within the HSD hypothesis.
		
We note that $k_{\mathrm{SM}}$ and $\xi_{LIV}$ do not follow the same behavior across the nuclei. As expected, the number of events necessary to constrain $\aof$ ($N_{\mathrm{ev}}$) is the lowest where the modulus of $\xi_{LIV}$ is the highest, although the relation is not linear and $N_{\mathrm{ev}}$ varies significantly from one nucleus to another. We also remark that applying the same procedure for an $\aof$ limit stronger by one order of magnitude than the most stringent current limit (\cite{NEMO-3-2019}) leads to an increase of two orders of magnitude of the needed number of events. This implies that in the near future, the DBD experiments will improve by some factor the best current upper limit of the $\aof$ coefficient.   
		
		\begin{table*}
			\begin{ruledtabular}
			\begin{tabular}{cccccc}
				Nucleus & Q-value (MeV)& $k_{\text{SM}}$ & $\xi_{\text{LIV}}\left(\mathrm{MeV}^{-1}\right)$ & $N\times 10^{-8}(|\aof| < 3\times 10^{-5}\mathrm{MeV})$ & $\varepsilon_{1}^{\mathrm{max}}$(MeV)\\
				$^{48}$Ca & 4.2681\cite{Qvalue-48Ca} & -0.7673 & -3.4931 & 8.4060 & 0.671\\
				$^{76}$Ge & 2.0391\cite{Qvalue-76Ge} & -0.5608 & -4.9831 & 4.4625 & 0.181\\
				$^{82}$Se$_{\text{SSD}}$ & 2.9979\cite{Qvalue-82Se} & -0.6585 & -4.3121 & 5.7670 & 0.197\\
				$^{100}$Mo$_{\mathrm{SSD}}$ & 3.0344\cite{Qvalue-100Mo} & -0.6690 & -4.2939 & 5.7932 & 0\\
				$^{110}$Pd & 2.0179\cite{Qvalue-110Pd} & -0.5788 & -5.0765 & 4.2760 & 0.120\\
				$^{116}$Cd & 2.8135\cite{Qvalue-116Cd} & -0.6726 & -4.3332 & 5.6808 & 0.192\\
				$^{130}$Te & 2.5275\cite{Qvalue-130Te} & -0.6514 & -4.6013 & 5.0779 & 0.220 \\
				$^{136}$Xe & 2.4587\cite{Qvalue-136Xe} & -0.6483 & -4.6828 & 4.9082 &  0.198\\
				$^{150}$Nd$_{\text{SSD}}$ & 3.3367\cite{Qvalue-150Nd} & -0.7218 & -4.1323 & 6.1258 & 0 \\
				$^{150}$Nd$_{\text{HSD}}$ & 3.3367 & -0.7357 & -3.9734 & 6.5869 & 0.375 \\
				
			\end{tabular}
			\end{ruledtabular}
			\caption{\label{tab:KXiNevts} $k_{\text{SM}}$ and $\xi_{\text{LIV}}$ computed as described in the text for all nuclei. $Q$-values used in calculations are also displayed. The fifth column contains the expected number of events needed to constrain $\aof$ to the current limit obtained from tritium decay \cite{KR-ARXIV}. The last column contains the position of the maxima of LIV single electron spectra}
		\end{table*}

\paragraph*{Conclusions.} We analyze the LIV effects on the single electron spectra, summed energy electron spectra, and angular correlation between electrons in $2\nu\beta\beta$ decay for all the experimentally interesting nuclei.
We derive the formulas of the LIV contributions to these spectra and angular correlation and provide theoretical predictions of them to be used for constraining the LIV coefficient $\aof$. Next, we analyze different signatures that could be probed in the DBD experiments. First, we confirm the overall effect of LIV to shift the single and summed energy electron spectra to higher electron energies for all the studied nuclei. Next, we highlight other LIV signatures that can be analyzed by comparing the electron and angular correlation spectra computed with and without LIV contributions and show that from this comparison, one can get information about the observability of the LIV effects in the current experimental statistics. Then, the alternative method of constraining $\aof$ from the measurement of the angular correlation coefficient is discussed. In this regard, we estimate the statistics that each of the DBD experiments, studying different nuclei, should reach to constrain  $\aof$ at the level of the current beta decay experiments. We hope that our work improves the theoretical support and further stimulates the search for LIV in DBD.  

\paragraph*{Acknowledgments}
	
	The figures for this article have been created using the SciDraw scientific figure preparation system \cite{SciDraw}.
	
	This work has been supported by the grants of the Romanian Ministry of Research, Innovation and Digitalization through the project PN19-030102-INCDFM and CNCS-UEFISCDI project no. 99/2021 within PN-III-P4-ID-PCE-2020-2374

%\newpage

\bibliography{thebibliography}
\bibliographystyle{apsrev4-2}

\end{document}